\documentclass[aps,pra,twocolumn,groupedaddress,showpacs]{revtex4-1}
\usepackage{graphicx}
\usepackage{amsmath}
\usepackage{textcomp}
\usepackage{mathrsfs}
\usepackage{amsfonts}

\newcommand{\GH}{Goos-H\"anchen }
\newcommand{\IF}{Imbert-Fedorov }

\renewcommand\Im{\mathrm{Im}}

\begin{document}
\title{Spatial \GH shift in photonic graphene}

\author{Simon Grosche}
\author{Alexander Szameit}
\author{Marco Ornigotti}
\email{marco.ornigotti@uni-jena.de}
\affiliation{
Institute of Applied Physics, Friedrich-Schiller-Universit\"at Jena, \\ Max-Wien-Platz 1, D-07743 Jena, Germany \\
}

\begin{abstract}
We present a theoretical study of the spatial \GH shift occurring at the interface between two different photonic graphene lattices. Numerical simulations of light propagation in such a wave guiding system are also presented to corroborate our findings.
\end{abstract}

\pacs{00.00, 20.00, 42.10}
\maketitle

\section{Introduction}
In the last decade graphene, a two dimensional honeycomb lattice of carbon atoms, has attracted a lot of interest, thanks to its intriguing properties \cite{katsnelson,ref2, ref3}. Its peculiar band structure and the existence of the so-called Dirac cones \cite{katsnelson}, for example, gives the possibility to use graphene as a model to observe QED-like effects such as Klein tunneling \cite{ref4}, Zitterbewegung \cite{ref5}, the anomalous quantum Hall effect \cite{ref6} and the appearance of a minimal conductivity that approaches the quantum limit $e^2/\hbar$ for vanishing charge density \cite{ref7}. Among the vast plethora of possible applications, graphene has proven to be a very interesting system for the observation of non specular reflection phenomena, such as the \GH (GH) \cite{ref19, ref20, ref21} and the \IF \cite{ref22, ref23} shifts. In particular, the occurrence of GH shift in graphene-based structures has been reported for light beams \cite{ref24, ref25} as well as for Dirac fermions \cite{ref26}.  A comprehensive review on beam shift phenomena and the \GH and \IF shifts can be found in \cite{refR3}. Although the first prediction of the GH shift dates back to 1947 \cite{refGH}, this field of research is still very active, and in the last decades a vast amount of literature has been produced on the subject, resulting in a better understanding of the underlying physics \cite{ref27, ref28, ref29} and the investigation of the effects of different field configurations \cite{ref30, ref31, ref32, ref33} and reflecting surfaces \cite{ref34, ref35, ref36, ref37}. In particular, a giant \GH shift has been observed to occur in various systems such as metamaterials \cite{ref39},  photonic crystals \cite{ref40}, complex crystals \cite{ref41}, and graphene-coated surfaces \cite{ref25}. Recently, moreover, the occurrence of GH shift has also been theoretically investigated for discrete periodic media such as optical waveguides, where a negative spatial shift has been also predicted  \cite{rechtsmanNegGH}. For the case of graphene, in particular, the occurrence of a giant GH shift is linked to its peculiar band structure, and in particular to the existence of Dirac points \cite{ref24, ref25}.  This distinctive dispersion relation, however, is not only a prerogative of graphene, but it is linked to the honeycomb lattice in which the carbon atoms are arranged \cite{katsnelson}. For this reason, in recent years graphene-like structures with honeycomb lattices have been implemented in other systems, such as for example electronic systems \cite{refR1, refR1a}, cold atoms \cite{refR1b} and photonic structures. In the latter case,  carbon atoms can for example be replaced by an array of waveduiges arranged in a honeycomb lattice, the so-called photonic graphene \cite{ref8, ref9, ref10, PTPhotGrpahPhysRevA.84.021806}, or by suitably designed photonic crystals \cite{ref12, ref13}. The properties induced by the presence of Dirac cones in these optical structures have been extensively studied both theoretically and experimentally, unraveling many interesting effects and dynamics such as conical diffraction \cite{ref8}, dynamical band collapse \cite{refR2}, topological protection of edge states \cite{ref15}, pseudo magnetic behaviour at optical frequencies \cite{ref16}, the first experimental realisation of an external field-free photonic topological insulator \cite{ref17} and the topology assisted dynamics of edge states \cite{ref18}, to name a few.

Motivated by all this, in this work we present a theoretical study of the occurrence of GH shift in photonic graphene. In particular, we foresee that the interplay between the honeycomb structure typical of graphene, and the discrete periodic nature of the waveguide array will play a central role in determining the properties of the observed GH shift. We in fact expect to observe a negative spatial shift given by the periodic nature of the system (in accordance with \cite{rechtsmanNegGH}), as well as a giant GH shift deriving from the graphene-like dispersion relation \cite{ref25}. 

This work is organised as follows: in Sect.\,2 we present a brief description of photonic graphene and we describe in detail the system used for observing the \GH shift. Sect.\,3 is then dedicated to the actual calculation of the spatial \GH shift, while the discussion of the results is given in Sect.\,4. Finally, conclusions are drawn in Sect.\,5.

\section{Model and Methods}
\subsection{Geometry of the system}
\begin{figure}[!t]
\begin{center}
	\includegraphics[width=0.5\textwidth]{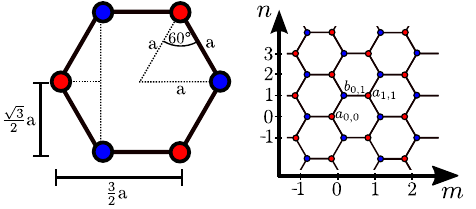}
\caption{On the left side, the Honeycomb geometry is shown. The waveguides are located at the vertices. On the right side, a mesh of the honeycomb structure is shown. The coordinates $m,n$ are illustrated. For all $a_{m,n}$, $m+n$ is chosen to be even (waveguides marked green). For all $b_{m,n}$, $m+n$ is chosen to be odd (waveguides marked yellow).}
\label{fig:honeycombGeo}
\end{center}
\end{figure}
We start our analysis by considering a monochromatic beam of light characterized by the frequency $\omega = c k$ (being $k$ the vacuum wave number), which propagates in a waveguide array arranged as a 2D honeycomb lattice, as shown in Fig.\,\ref{fig:honeycombGeo}.
Light propagation in such a system is well described by the optical Schr\"odinger equation \cite{waveguideDiscreteOptics}
\begin{equation}
	\label{eq:schroedinger}
	i\lambdabar\frac{\partial E}{\partial z}=-\frac{\lambdabar^2}{2n_0}\nabla_{\perp}^2E+\Delta n(x,y) E,
\end{equation}
where $E=E(x,y,z)$ is the amplitude of the electric field envelope, $n_0$ is the bulk refractive index, $\lambdabar=1/k$ is the reduced wavelength and $\Delta n(x,y)=n(x,y)-n_0$ is the refractive index contrast.
By expanding the field amplitude $E$ onto the modes of the individual waveguides and considering only nearest neighbour coupling, the so-called coupled mode equations, i.e., the tight-binding model, are obtained. In the case of photonic graphene depicted in Fig.\,\ref{fig:honeycombGeo}, this brings to the following set of coupled mode equations \cite{PTPhotGrpahPhysRevA.84.021806}
\begin{subequations}\label{eq2}
\begin{align}
	(\mathrm{i}\partial_z + \delta\beta) a_{m,n}+\kappa (b_{m-1,n}+b_{m,n-1}+b_{m,n+1})&=0 ,\\
	(\mathrm{i}\partial_z + \delta\beta) b_{m,n}+\kappa(a_{m+1,n}+a_{m,n-1}+a_{m,n+1})&=0 ,
\end{align}
\end{subequations}
where $a_{m,n}$ and $b_{m,n}$ are the complex amplitudes in the $(m,n)$-th waveguide of the honeycomb lattice, $\delta\beta$ is the propagation constant mismatch between the sublattices and $\kappa$ is the coupling constant between the individual waveguides. The propagation constant mismatch $\delta\beta$ can be set equal to zero, as it only contributes in an overall, inessential phase factor \cite{snyder1983optical}.

\begin{figure}[t!]
\begin{center}
	\includegraphics[width=0.5\textwidth]{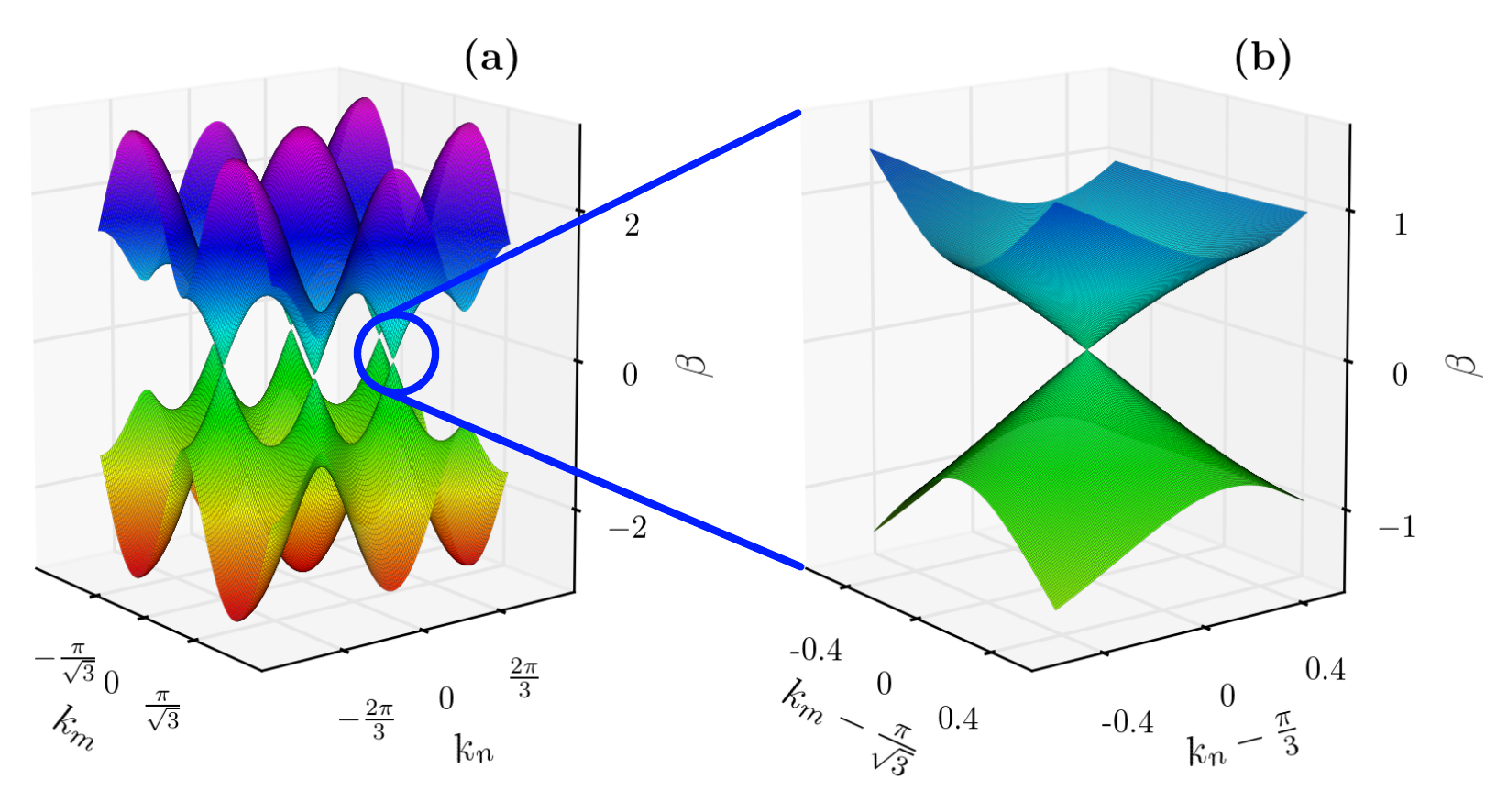}
	\caption{(a) Dispersion relation of photonic graphene. The upper and the lower band touch each other in the Dirac points. Near these points, the dispersion relation is linear (b). Here, $\kappa=1$ and $\delta\beta=0$ have been used.}
	\label{fig:dispersionGraphene}
\end{center}
\end{figure}

Due to the periodicity of the system, assumed to be infinitely extended in the transverse direction, solutions of Eqs. (2) can be searched in the form of Bloch waves, namely
\begin{equation}\label{BlochAnsatz}
	\left(\begin{array}{c}
	a_{m,n}\\b_{m,n}
	\end{array}\right)=
	\left(\begin{array}{c}
	A\\B
	\end{array}\right) 
	\mathrm{e}^{\mathrm{i}(\beta z + \sqrt{3} k_m m +k_n n)},
\end{equation}
where $\beta$ is the longitudinal propagation constant and $ k_m$ and $k_n$ are the transversal wave numbers.  Inserting the ansatz given by Eq.\eqref{BlochAnsatz} into Eqs.\eqref{eq2} yields to the following dispersion relation
\begin{equation}
	\label{eq:eigenvalue}
	\beta_\pm =\delta\beta \pm \kappa \sqrt{1+4\cos^2 k_n+4\cos k_n\cos\sqrt{3} k_m}\,\,.
\end{equation}
The explicit form of the above dispersion relation is depicted in Fig.\,\ref{fig:dispersionGraphene}. As it can be seen, the two bands touch each other in six points, the so called Dirac points. Near these points, as it appears clear from  Fig.\,\ref{fig:dispersionGraphene}(b), the dispersion relation becomes linear. 
From Eqs.\eqref{eq2} and by using the dispersion relation given in Eq.\eqref{eq:eigenvalue}, the explicit expression of the Bloch modes associated to this lattice can be written as follows:
\begin{equation}
	\mathbf{v}_\pm =\frac{1}{N}
	\left(\begin{array}{cc}
 	\beta_\pm-\delta\beta\\
	\kappa\mathrm{e}^{\mathrm{i}\sqrt{3} k_m}+2\kappa\cos k_n
	\end{array}\right)
	,
	\label{eq:eigenvector}
\end{equation}
where  $N$ is a suitable normalisation constant. 
The plus signs in Eqs.\eqref{eq:eigenvalue} and \eqref{eq:eigenvector} correspond to the case in which the amplitudes in all waveguides are in phase. In this situation, light finds itself in the upper band. If, on the contrary, the minus sign is assumed, there is a phase difference between adjacent waveguides. This corresponds to light finding itself in the lower band. In the remaining of this manuscript, we will only consider the case in which the upper band is occupied. Therefore, the notation $\beta\equiv\beta_{+}$ and $\mathbf{v}( k_m,\kappa, \delta\beta)\equiv\mathbf{v}_{+}$ is used.
\begin{figure}[t!]
\begin{center}
	\includegraphics[width=0.5\textwidth]{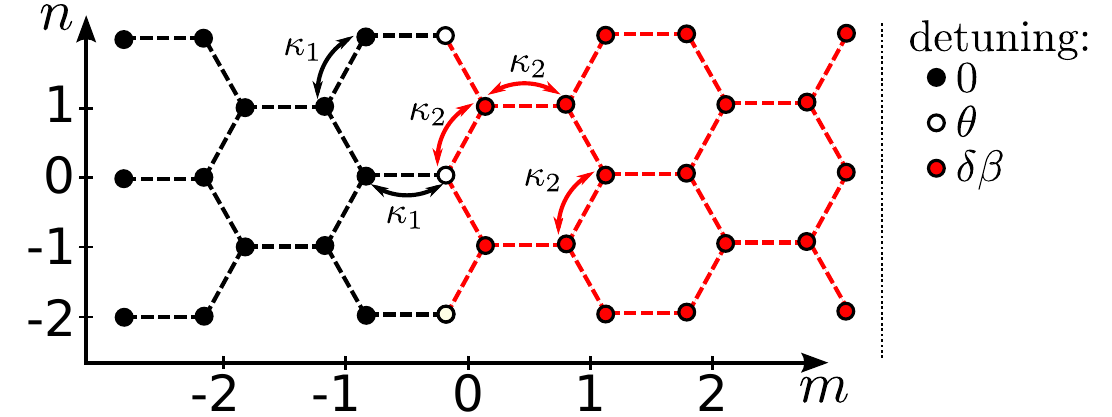}
	\caption{Schematic representation of the used interface in the waveguide array. The interface is induced by different coupling constants ($\kappa_1$, $\kappa_2$) and a change of the detuning ($0$, $\theta$, $\delta\beta$).}
	\label{fig:InterfaceBearded}
\end{center}
\end{figure}

\subsection{Interface and total reflection}
\label{sec:interfaceTotalReflecion}
To observe the \GH shift, an interface of some sort is needed. Here, we realize such an interface by inserting a defect line at an arbitrary position inside the lattice, as shown in Fig.\,\ref{fig:InterfaceBearded}. Moreover, we assume that the two sides of the lattice are infinitely extended, and they are characterized by the coupling constants $\kappa_1$ and $\kappa_2$, respectively. Furthermore, the detuning on the left side of the interface is set to zero, whereas the detuning on the right side is $\delta\beta$. The waveguides composing the interface may or may not have an extra detuning. If the interface detuning is called $\theta$, two scenarios are possible, namely $\theta=0$ (the interface waveguides have the same detuning as the left part of the lattice) and $\theta=\delta\beta$ (the interface waveguides have the same detuning as the right part of the lattice) \cite{fresnelpaper}.

The dispersion relations of both parts of the system can now be calculated using Eq.\eqref{eq:eigenvalue}. In order to maintain translational invariance of the system in the $z$-direction, the propagation constants of both the left and the right side must be equal at any time,  i.e., the two dispersion relations $\beta_\mathrm{left}(k)$ and $\beta_\mathrm{right}(k)$ must satisfy $\beta_\mathrm{left}(k) = \beta_\mathrm{right}(k)$. If the interface is assumed to be infinitely extended along the vertical direction (see Fig.\,\ref{fig:InterfaceBearded}), then also $k_n$ must be conserved across the interface. This leaves $ k_m^{(i)}$ as the only degrees of freedom. Here, we use the upper index to identify the left ($i=1$) and right ($i=2$) side, respectively. Using Eq.\eqref{eq:eigenvalue}, the constraint $\beta_\mathrm{left}=\beta_\mathrm{right}$ for the two propagation constants can then be rewritten as
\begin{eqnarray}
\beta_{\pm}(0, k_m^{(1)}, \kappa_1) = \beta_{\pm}(\delta\beta, k_m^{(2)}, \kappa_2)
\label{eq:EqualDisp}
\end{eqnarray}
where we implicitly used the shorthand notation $\beta_{\pm}=\beta_{\pm}(\delta\beta,k_m,\kappa)$. A closer inspection of the above equation shows that we can use two parameters to describe the refraction in such a system, namely the ratio $\kappa_1/\kappa_2$ between the coupling constants of the two lattices, and the normalized propagation mismatch $\delta\beta/\kappa_2$. This is in accordance with the results obtained for the one dimensional case \cite{fresnelpaper}. If we consider the band structure of the two lattices, in fact, we can make the same observation as in the one dimensional case: if the two dispersion relations do not overlap, no real solution for $ k_m^{(2)}$ exists. This means that a wave, say, initially travelling in the left side of the interface cannot penetrate into the right side, because it becomes evanescent. In this case, therefore, total reflection will occur. From Eq.\eqref{eq:EqualDisp}, it follows that a sufficient condition for this to take place is given by

\begin{equation}
	|\delta\beta| \ge 3 (\kappa_1+\kappa_2).
	\label{eq:condTotal}
\end{equation}
\begin{figure*}[!t]
\begin{center}
\includegraphics[width=\textwidth]{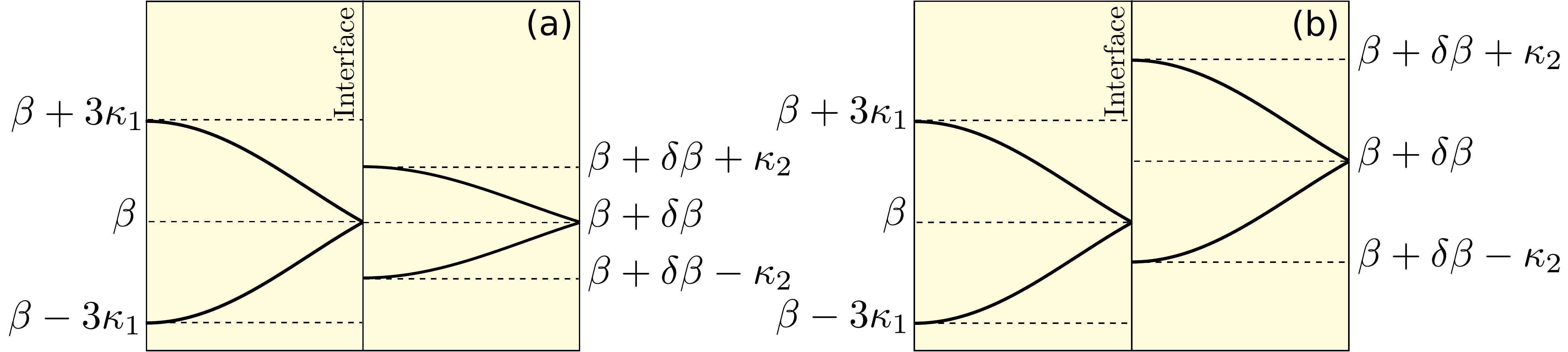}
\caption{Comparison of the dispersion relation (with $ k_m=0$) on both sides of the interface. In (a), the dispersion on the right side is scaled due to the choice $\kappa_1 < \kappa_2$, $\delta\beta=0$. In (b) it is only shifted by choosing $\kappa_1 = \kappa_2$, $\delta\beta > 0$.}
\label{fig:TwoDim-DispersionCompare}
\end{center}
\end{figure*}
\subsection{Reflection coefficients}
\begin{figure*}[!t]
\begin{center}
\includegraphics[width=\textwidth]{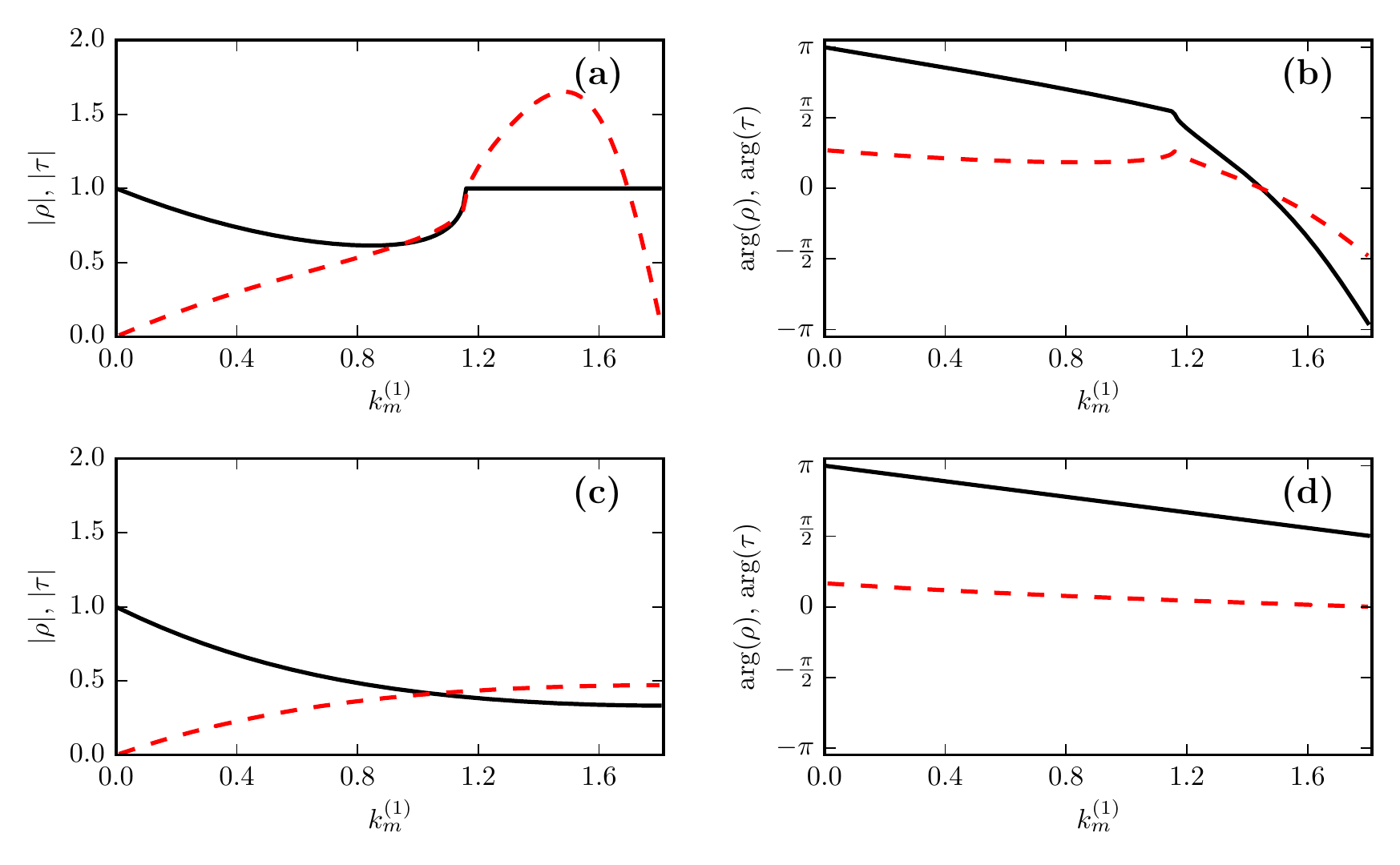}
\caption{Modulus (left column) and phase (right column) of the reflection (black, solid) and transmission (red, dashed) coefficients with parameters  $\delta\beta=\theta=0$, $\kappa_1=1$, $\kappa_2=2$. The top row shows the behaviour of $\rho$ and $\tau$ away from the Dirac point (i.e., $k_n=0.4$), while in the bottom row, the reflection and transmission coefficients in correspondence to the Dirac point (i.e., $k_n=\frac{\pi}{3}$) are depicted.}
\label{fig:reftransBeraded}
\end{center}
\end{figure*}

\begin{figure*}[!t]
\begin{center}
\includegraphics[width=\textwidth]{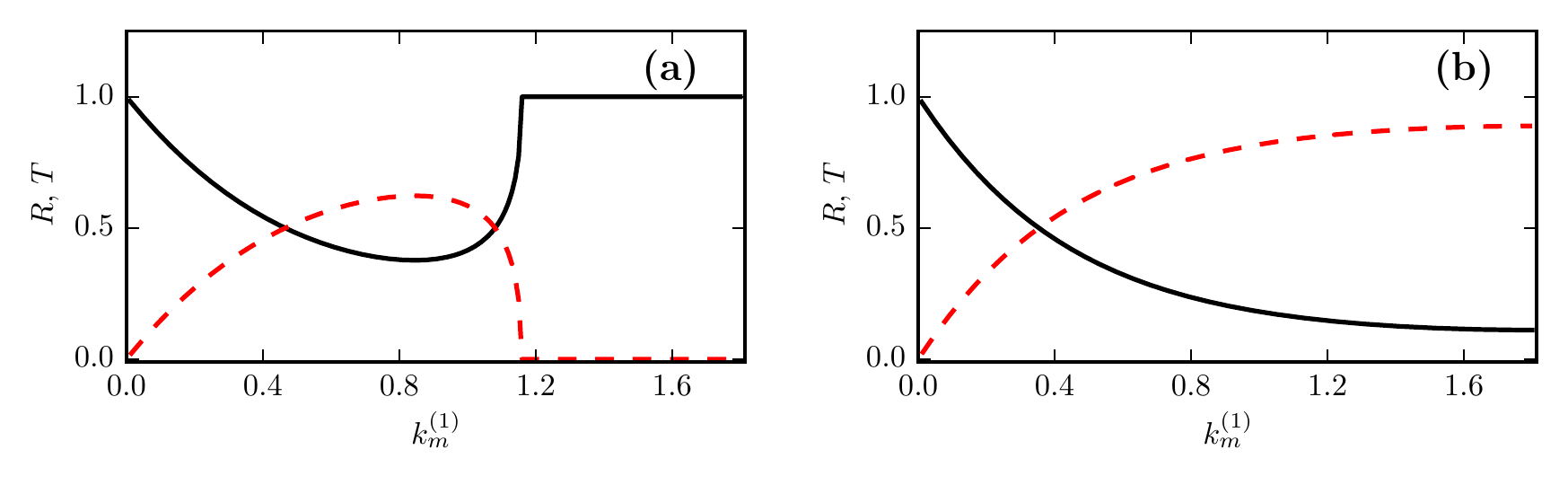}
\caption{Reflection (black, solid) an transmission (red, dashed) coefficients of the power with parameters  $\delta\beta=\theta=0$, $\kappa_1=1$, $\kappa_2=2$. The left panel shows the case where $k_n$ is chosen away from the Dirac point (i.e., $k_n=0.4$), while the right panel depicts the case when $k_n$ has been chosen in correspondence of the Dirac point(i.e., $k_n=\frac{\pi}{3}$).}
\label{fig:REFtransBeraded}
\end{center}
\end{figure*}

The model described above explains the reflection only qualitatively. There is no prediction, in fact, about the amplitude or the phase of the reflected light.
These quantities can be then explicitly calculated by employing the same approach used in Ref.\,\cite{fresnelpaper}, where detailed calculations of the reflection and transmission coefficients of a discrete system are carried out for a one dimensional waveguide array. The purpose of this section is therefore to generalize those results to the case of photonic graphene, with particular attention to the case depicted in Fig.\,\ref{fig:InterfaceBearded}. In this case, the propagation of light in the vicinity of such an interface can be described by the following set of three differential equations:
\begin{widetext}
\begin{subequations}\label{eq:dglInterfaceAll}
\begin{align}
	\mathrm{i}\partial_z b_{-1,2n}+ \kappa_1 a_{0,2n}+ \kappa_1 a_{-1,2n+1}+\kappa_1 a_{-1,2n-1}&=0 ,\label{eq:dglInterfaceLeft}\\
	(\mathrm{i}\partial_z + \theta) a_{0,2n} + \kappa_1 b_{-1,2n}+ \kappa_2 b_{0,2n+1}+ \kappa_2 b_{0,2n-1}&=0 ,\label{eq:dglInterfaceMid}\\
	(\mathrm{i}\partial_z + \delta\beta) b_{0,2n+1}+ \kappa_2 a_{1,2n+1}+ \kappa_2 a_{0,2n+2}+ \kappa_2 a_{0,2n}&=0\label{eq:dglInterfaceRight}.
\end{align}
\end{subequations}
\end{widetext}
Equations\eqref{eq:dglInterfaceLeft} and \eqref{eq:dglInterfaceRight} describe the propagation of light in the left and right lattice, respectively, while Eq.\eqref{eq:dglInterfaceMid}  ensures that the waveguides at the interface couple correctly to the two neighbouring lattices. The reflection and transmission coefficients $\rho$ and $\tau$ can be then calculated by assuming that the solution to the above equations can be written in the following form:
\begin{widetext}
\begin{equation}\label{eq:ansatzInterface}
	\left(\hspace*{0.5mm}\begin{array}{c}
	a_{m,n}\\
	b_{m,n}
	\end{array}\hspace*{0.5mm}\right) = \left\lbrace \hspace*{0.5mm}\begin{array}{ll}
	 \left[\mathbf{v}( k_m^{(1)},\kappa_1,0) +\rho\,  \mathrm{e}^{-\mathrm{i}\phi_m}\mathbf{v}(- k_m^{(1)},\kappa_1,0)\right]\Psi^{(1)},&\mathrm{if}\, m<0, \\
	
	 \tau'\,  \mathbf{v}( k_m^{(3)},\kappa_3,\theta)\Psi^{(2)}
	,&\mathrm{if}\, m=0 ,\,n\,\mathrm{even},\\
	
	\tau\, \mathbf{v}( k_m^{(2)},\kappa_2,\delta\beta)\Psi^{(3)} 
	,&\mathrm{otherwise},
	
	\end{array} \right.
\end{equation}
\end{widetext}
where $\tau'$ is the transmission coefficient into the defect (i.e., the interface waveguide),  $\phi_m=2 \sqrt{3} k_m^{(1)} m$ and $\Psi^{(k)}=\exp{\left[\mathrm{i}(\beta z +k_n n + \sqrt{3} k_m^{(k)} m)\right]}$, with $k\in \left\lbrace1,2,3\right\rbrace$. One should note that the wavenumber of the reflected Bloch wave is equal to $- k_m^{(1)}$, because it is reflected at the interface and the direction of the propagation changes according to the law of specular reflection \cite{bornWolf}.  
Solving Eqs.\eqref{eq:dglInterfaceLeft}--\eqref{eq:dglInterfaceRight} with the help of Eq.\eqref{eq:ansatzInterface} gives the following result for the reflection coefficient $\rho$:
\begin{equation}
\rho=-\frac{f_1( k_m^{(1)}, k_m^{(2)})-f_2( k_m^{(1)}, k_m^{(2)})}{f_1(- k_m^{(1)}, k_m^{(2)})-f_2(- k_m^{(1)}, k_m^{(2)})}	,
\end{equation}
where 
\begin{subequations}
\begin{align}
	\mathrm{f}_1(x,y)&=-2\kappa_2v_1(x,\kappa_1,0) \Big[\cos k_n\, v_2(y,\kappa_2,\delta\beta)\nonumber\\
	&-\Delta\beta \,v_1(y,\kappa_2,\delta\beta)\Big],\\
	\mathrm{f}_2(x,y)&=\kappa_1\mathrm{e}^{-\mathrm{i}\sqrt{3}x}\, v_2(x,\kappa_1,0) \,v_1(y,\kappa_2,\delta\beta),
\end{align}
\end{subequations}
where $\Delta\beta=(\beta-\theta)/2\kappa_2$ and  $v_1$($v_2$) is the first (second) component of the eigenvector $\mathbf{v}_+$ as given by Eq.\eqref{eq:eigenvector}. The calculation of the transmission coefficients $\tau$ and $\tau'$ follows straightforwardly.
The reflection and transmission coefficients are shown in Fig.\,\ref{fig:reftransBeraded}. At first glance, these results can appear counterintuitive, since the transmission coefficient $|\tau|$ is nonzero even when $|\rho|=1$ [see Fig.\,\ref{fig:reftransBeraded} (a)]. Similar findings have been made in Ref.\,\cite{fresnelpaper}, where this point is discussed in detail for the case of a 1D array of waveguides. In the present case, a careful calculation of the reflectivity $R$ and transitivity $T$ yields
\begin{eqnarray}
\label{eq:ReflTransPower}
R&=|\rho|^2,\\
T&=|\tau|^2 \, \frac{\kappa_2 \, \Im[ v_1^* (k_m^{(2)},\kappa_2,\delta\beta) \, v_2(k_m^{(2)},\kappa_2,\delta\beta)]}{ \kappa_1 \,\Im[ v_1^*(k_m^{(1)},\kappa_1,0) \, v_2(k_m^{(1)},\kappa_1,0) ]} \, ,
\end{eqnarray}
with $R+T=1$.
Both quantities are shown in Fig.\,\ref{fig:REFtransBeraded} for the two cases from Fig.\,\ref{fig:reftransBeraded}. One can see that the equation $R+T=1$ is fulfilled.

\section{Spatial \GH shift}
\label{sec:GHShiftTheory}
To observe the spatial \GH shift in case of an dielectric interface, total internal reflection must occur. This corresponds to having a reflection coefficient which can be written as a pure phase factor $\rho=\exp(\mathrm{i}\phi)$ \cite{GHfirst}. For a Gaussian beam impinging onto a dielectric interface, the explicit expression of the spatial \GH shift is given by \cite{Jopt_Ornigotti,rechtsmanNegGH}
\begin{equation}\label{GHS}
\delta_{GH}=-\frac{\partial \phi}{\partial \beta'},
\end{equation}
where $\beta'$ is the longitudinal component of the wave vector. $\delta_{GH}$ is the shift along the direction of propagation. In a real experiment, however, this parameter is not accessible, as it is possible to acquire information on the evolution of light by either monitoring the fluorescence caused by the light during propagation \cite{fluorescence}, i.e., to observe the light pattern along $z$ from the top facet of the sample, or to monitor directly the intensity distribution at the exit facet of the sample. For this reason, therefore, it is better to consider the projection of Eq.\eqref{GHS} along the $z$ direction, namely
\begin{equation}
\Delta_{GH}=-\frac{\partial \phi}{\partial \beta},
\end{equation}
which is a parameter that can be accessed experimentally. The above equation, moreover, can be rewritten in a more useful way by noticing that the phase $\phi$ of the reflection coefficient $\rho$ is a function of the transverse momentum $ k_m$ and therefore the following equality holds:
\begin{equation}\label{eq:GHshihftbm}
	\Delta_{GH}=-\frac{\partial\phi}{\partial\beta}=-\frac{\partial  k_m}{\partial \beta} \frac{\partial \phi}{\partial  k_m} .	
\end{equation}
Notice, moreover, that the quantity $\partial k_m/\partial\beta$ is nothing but the inverse of the group velocity along the $m-$direction, and can be calculated analytically from Eq.\eqref{eq:eigenvalue}.
\section{Results and Discussion}
\begin{figure}[t!]
	\includegraphics[width=0.5\textwidth]{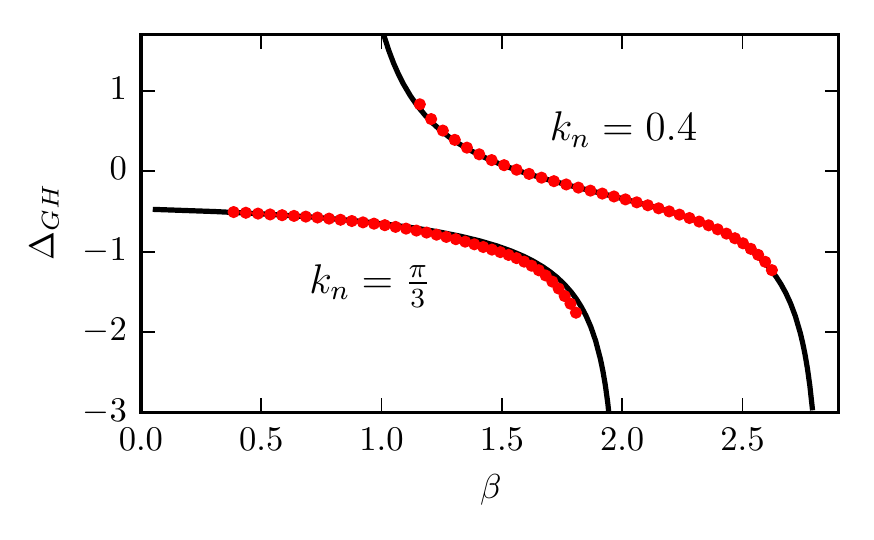}
	\caption{Dimensionless \GH shift as obtained with analytical theory given by Eq. \eqref{eq:GHshihftbm} (solid black line) and numerical simulations obtained by integrating Eqs. (2) (red dots). The parameters used for obtaining both graphs are $\kappa_1=\kappa_2=1$ and $\delta\beta=\theta=3(\kappa_1+\kappa_2)$ in order to ensure total internal reflection. Two cases, $k_n=0.4$ and $k_n=\pi/3$, are shown.}
	\label{fig:GHBeradedTheoSim}
\end{figure}

The spatial \GH shift for a Gaussian beam totally reflecting from the side edge of two juxtaposed honeycomb lattices is shown in Fig.\,\ref{fig:GHBeradedTheoSim}. Here, we compare the analytical calculations (black, solid line) obtained from Eq.\eqref{eq:GHshihftbm} with the numerical results (red dots) obtained by simulating the propagation of light in such a structure. Moreover, we consider two cases, namely the case $k_n=0.4$, corresponding to being away from the Dirac point, and $k_n=\pi/3$ corresponding to being close to a Dirac point.
For the first case, namely $k_n=0.4$, the \GH shift may be either positive or negative, depending on the chosen value of $\beta$. Away from the Dirac points, in fact, the \GH shift is comparable to the one dimensional system that is investigated in \cite{rechtsmanNegGH}. This is reasonable because the shapes of both dispersion relations are not too different and in particular both dispersion relation present a non-zero effective mass. The effective mass, which is proportional to the second derivative of the band structure, is in fact negative on the right side of the interface. For small values of $\beta$, however, it is also negative on the left side as well. Therefore, a positive shift is expected. For larger $\beta$, the effective mass on the left side of the interface becomes positive, therefore allowing a negative \GH shift.

For $k_n=\frac{\pi}{3}$, i.e., close to a Dirac point, the shift is always negative. For large $\beta$, the argument from the previous case, $k_n=0.4$, still holds as the shapes of the dispersion relations are comparable. For small $\beta$, i.e. $\beta \rightarrow 0$, a Dirac point is reached and the dispersion relation is approximately linear. Consequently, the effective mass is zero and the shift stays negative.
In order to prove our theoretical results, we have performed a numerical simulation of the propagation of light in a system consisting of two juxtaposed honeycomb lattices separated by an interface constituted by a line of waveguides with different propagation constant [see Fig.\,\ref{fig:InterfaceBearded}] using the zvode algorithm \cite{zvode}. Periodic boundary conditions have been used in the $n$-direction and the waveguides were excited with a broad Gaussian beam profile 

\begin{figure*}[t!]
\begin{center}
\includegraphics[width=\textwidth]{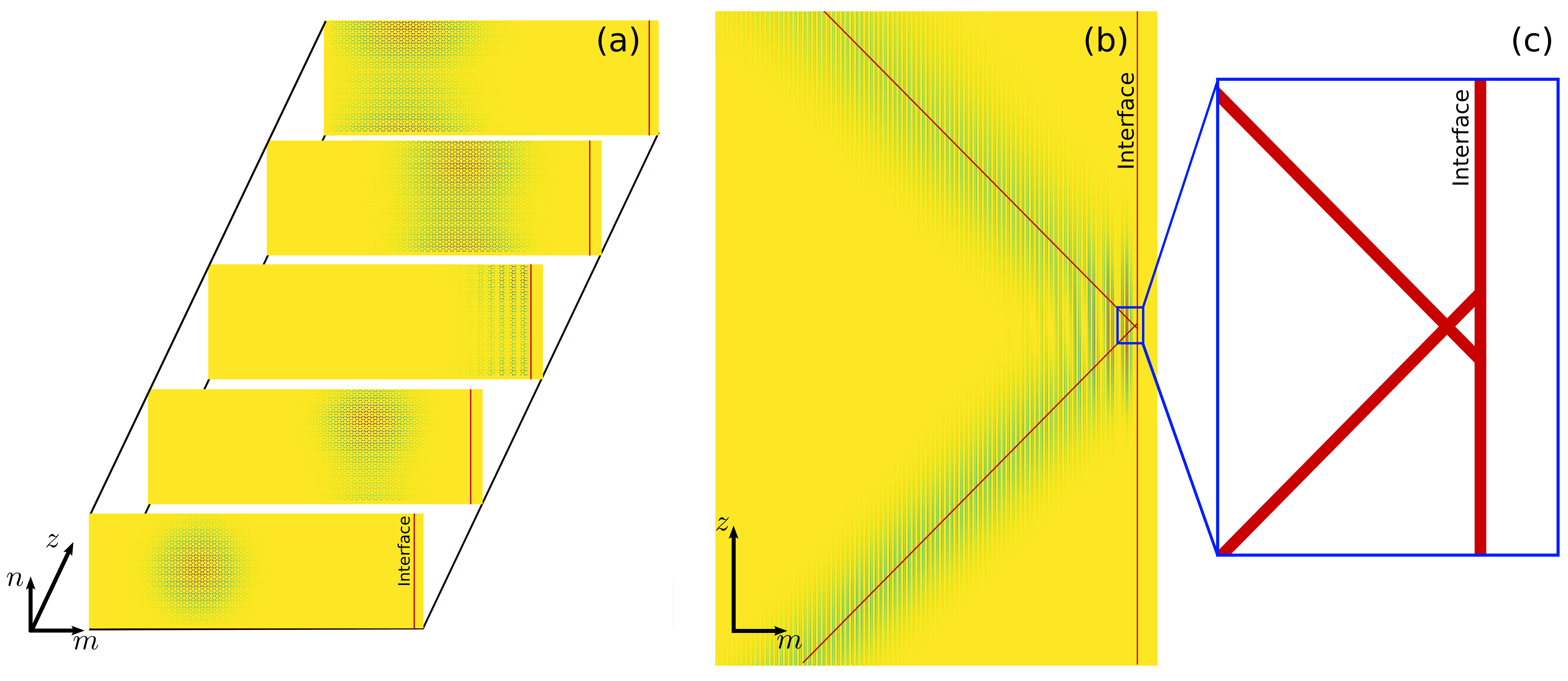}
\caption{Light evolution in the photonic graphene lattice showing a negative \GH shift. Various snapshots are shown in (a), whereas a view from above is shown in (b). The interface is marked with a solid white line. In (a), the initial condition, i.\,e. a wide Gaussian profile, can be seen in the first picture and after some propagation, the interaction with the interface can be seen in the third picture. Here, no light propagates to the right part. In (b), the negative \GH shift can be seen considering the extrapolated light rays of geometric-optics which intersect in front of the interface.}
\label{fig:numericalSim}
\end{center}
\end{figure*}

The results of our numerical simulations are shown in Fig.\,\ref{fig:numericalSim}\,(a) for some distances $z$, while in Fig.\,\ref{fig:numericalSim}\,(b), the extrapolated light rays of the geometric-optics approximation are shown. The light rays of the incoming and outgoing beam do not intersect on the interface but in front of it. A closer inspection of the ray optics extrapolation shows that the point of reflection is shifted towards negative valued with respect to the geometrical optics reflection point, as it can be seen from the inset of Fig.\,\ref{fig:numericalSim}\,(b). To corroborate this fact, we have numerically integrated Eqs. (2) and then computed the \GH shift for various values of $ k_m$ and compared these results with the one obtained analytically from Eq.\eqref{eq:GHshihftbm}. The result of this comparison can be seen in Fig.\,\ref{fig:GHBeradedTheoSim}. As it can be seen, the numerical simulation is in very good agreement with our theoretical predictions.

\section{Conclusions}
In this work, we have presented a theoretical analysis of the occurrence of the \GH shift at the interface of two juxtaposed honeycomb lattices. Our findings show that in the vicinity of a Dirac point, the \GH shift remains always negative. Away from the Dirac point, instead, our results agree with the one reported in \cite{rechtsmanNegGH} for the case of a 1D periodic structure. Numerical simulations of the reflection of a Gaussian beam at the interface of such a structure has also been presented to corroborate our theoretical results. 

\section*{Acknowledgements}
The authors acknowledge funding by the German Research Foundation (grants SZ 276/9-1, SZ 276/7-1 and BL 574/13-1).

\end{document}